\documentclass[fleqn,twoside]{article}
\usepackage{espcrc2}
\usepackage{graphicx}

\newcommand{\la}{\langle}
\newcommand{\ra}{\rangle}
\newcommand{\op}{{\cal O}}
\newlength{\minitwocolumn}
\newcommand{\msbar}{{\overline {\rm MS}}}
\newcommand{\KEK}
{High Energy Accelerator Research Organization (KEK),
 Tsukuba, 305-0801, Japan}

\newcommand{\Tsukuba}
{Institute of Physics, University of Tsukuba,
 Tsukuba, 305-8571, Japan}

\newcommand{\ICRR}
{Institute for Cosmic Ray Research, University of Tokyo,
 Kashiwa, 277-8582, Japan}

\newcommand{\RCCP}
{Center for Computational Physics, University of Tsukuba,
 Tsukuba, 305-8577, Japan}

\newcommand{\Hiroshima}
{Department of Physics, Hiroshima University,
 Higashi-Hiroshima, 739-8526, Japan}

\newcommand{\YITP}
{Yukawa Institute for Theoretical Physics, Kyoto University,
 Kyoto, 606-8502, Japan}

\title{
Continuum limit of proton decay matrix elements in quenched
lattice QCD
\thanks{Talk presented by N.~Tsutsui}
}

\author{
CP-PACS and JLQCD Collaborations:
N.~Tsutsui\address{\KEK},
S.~Aoki\address{\Tsukuba},
M.~Fukugita\address{\ICRR},
S.~Hashimoto$^{\rm a}$,
K-I.~Ishikawa$^{\rm b,}$\address{\RCCP},
T.~Ishikawa$^{\rm d}$,
N.~Ishizuka$^{\rm b,d}$,
Y.~Iwasaki$^{\rm b}$,
K.~Kanaya$^{\rm b}$,
T.~Kaneko$^{\rm a}$,
Y.~Kuramashi$^{\rm a}$,
M.~Okawa\address{\Hiroshima},
T.~Onogi\address{\YITP},
N.~Taniguchi$^{\rm b}$,
A.~Ukawa$^{\rm b,d}$,
and
T.~Yoshi\'e$^{\rm b,d}$
}
       
\begin{document}
\begin{abstract}
We present a lattice QCD calculation of the parameters
$\alpha$ and $\beta$ which are necessary in
the theoretical estimation of the proton lifetime
in grand unified theories (GUTs) using chiral lagrangian approach.
The simulation is carried out using the Wilson quark action
at three gauge coupling constants in the quenched approximation.
We obtain
$|$$\alpha$(2GeV)$|$=0.0091(08)($^{+10}_{-19}$)GeV$^3$
and
$|$$\beta$(2GeV)$|$=0.0098(08)($^{+10}_{-20})$GeV$^3$
in the continuum limit
where the first error is statistical and the second one
is due to scale setting.
\end{abstract}
\maketitle
\section{Introduction}

Proton decay is one of the most important predictions
of grand unified theories (GUTs)
and a lot of experimental effort has been devoted
to detect it.
Up to now, however, no such decay process have been observed, and 
the lower bound of the proton lifetime 
has been pushed up to $O(10^{32})$ years 
by the Super-Kamiokande experiment.

To constrain the parameter space of GUTs by the proton lifetime,
we need a reliable estimation of the hadronic matrix elements
$\la PS | \op | N \ra$, 
where $PS$ and $N$ represents pseudoscaler meson and nucleon
respectively and $\op$ is a three-quark operator.
In principle, lattice QCD allows a precise determination 
of the matrix elements from the first principles.
In previous work, we made a GUT-model-independent calculation 
of the matrix elements for all dimension-six three-quark operators
using the Wilson quark action and the plaquette gauge action
in the quenched approximation\cite{Aoki:1999tw}.

There remain two major systematic errors in the previous calculation:
scaling violation and quenching effects.
As a first step to reduce the systematic errors,
this work is devoted to study the scaling violation effects.
We leave incorporation of the dynamical quark effects to future work.

\section{Method}

With the aid of chiral perturbation theory\cite{Claudson:1981gh},
the proton decay matrix elements
$\la PS | \op | N \ra$ are related to 
the so-called $\alpha$ and $\beta$ parameters, which are
defined by
\begin{eqnarray}
 \la 0 |
 \epsilon_{ijk}
 (u^i CP_R d^j) P_L u^k
 | p(\vec{k}=\vec{0}) \ra &=&
 \alpha P_L u_p,\\
 \la 0 |
 \epsilon_{ijk}
 (u^i CP_L d^j) P_L u^k
 | p(\vec{k}=\vec{0}) \ra &=&
 \beta P_L u_p,
\end{eqnarray}
where $u_p$ denotes the proton spinor.
Calculation of these parameters is much simpler than
the direct calculation of the matrix elements $\la PS | \op | N \ra$:
the former is obtained from two-point functions with zero 
spatial momentum projection, while the latter
requires three-point functions with finite spatial momenta.
It should be noted that
our previous study showed that
the matrix elements obtained from the three-point functions
are roughly comparable with
the tree-level predictions of
the chiral lagrangian
with the $\alpha$ and $\beta$ parameters determined on the lattice.

\begin{table}[t]
\caption{
 Simulation parameters. The lattice spacing $a$ is determined from $m_\rho$.
}
\label{simulation_parameters}
\begin{tabular}{cccc}
\hline
 $\beta$ & $L^3\times T$ & $a^{-1}$[GeV] & \#conf. (Ref.~\protect{\cite{Aoki:2002fd}})\\
\hline
 5.90 & 32$^3\times$56 & 1.934(16) & 300 (800)\\
 6.10 & 40$^3\times$70 & 2.540(22) & 200 (600)\\
 6.25 & 48$^3\times$84 & 3.071(34) & 140 (420)\\
\hline
\end{tabular}
\end{table}

The lattice operators relevant for the $\alpha$ and
$\beta$ parameters are renormalized as
\[
 \op_{R/L,L}^{\rm{cont}}(\mu) =
 Z(\alpha_s,\mu a) \op_{R/L,L}^{\rm{latt}}(a)
\]
\begin{equation}
 +\frac{\alpha_s}{4\pi} Z_{\rm{mix}} \op_{L/R,L}^{\rm{latt}}(a)
 \mp\frac{\alpha_s}{4\pi} Z^{\prime}_{\rm{mix}} \op_{\gamma_{\mu}L}^{\rm{latt}}(a),
\end{equation}
where
\begin{equation}
 \op_{R/L,L} =
 \epsilon_{ijk}
 (u^i CP_{R/L} d^j) P_L u^k,
\end{equation}
and the additional mixing operator is
\begin{equation}
 \op_{\gamma_{\mu}L} =
 \epsilon_{ijk}
 (u^i C\gamma_{\mu}\gamma_5 d^j) P_L\gamma_{\mu} u^k.
\end{equation}
The renormalization constants $Z$, $Z_{\rm{mix}}$, and
$Z^{\prime}_{\rm{mix}}$ are calculated perturbatively at
one-loop level\cite{Aoki:1999tw}. The continuum operators are 
defined in NDR (naive dimensional regularization) 
with the $\msbar$ subtraction scheme. 

To obtain the matrix elements, we compose a ratio
\begin{equation}
 R(t)=
 \frac{
 \sum_{\vec{x}}
 \la \op_{R/L,L}(\vec{x},t)
 \bar{J}^{\prime}_{p,s}(0)\ra
 }
 {
 \sum_{\vec{x}}\la J_{p,s}(\vec{x},t)\bar{J}^{\prime}_{p,s}(0)\ra
 }
 \sqrt{Z_p},
\end{equation}
where $\sqrt{Z_p}$ is defined by
$
 \la 0 | J_{p,s}({\vec 0},0) | p^{(s^\prime)}({\vec 0}) \ra =
 \sqrt{Z_p} u_s^{(s^\prime)}.
$
This factor is estimated from the
proton correlator with local source and local sink.
It is well known that a precise determination of $\sqrt{Z_p}$ is
rather hard
because of large statistical fluctuations of the local-local correlator.
On the other hand, the ratio of two-point functions can be calculated using 
the smear-local proton correlator
whose statistical fluctuation is fairly small compared to the
local-local correlator.

This situation leads us to the following strategy:
(i) For $\sqrt{Z_p}$ we resort to the result obtained
from high statistical calculation of the quenched light hadron spectrum
performed by the CP-PACS collaboration\cite{Aoki:2002fd}.
(ii) Employing the same parameters
as the CP-PACS calculation, we make a new simulation 
to estimate the ratio of two-point functions 
including the mixing operators,
which were not incorporated in the previous CP-PACS simulation.
After combining the results of $\sqrt{Z_p}$ and 
the ratio of two-point functions we can 
achieve  a few \% level of statistical accuracy  
for the matrix elements at each $\beta$ and hopping parameter.

\begin{figure}[t]
  \includegraphics[width=\minitwocolumn,height=5.2cm]{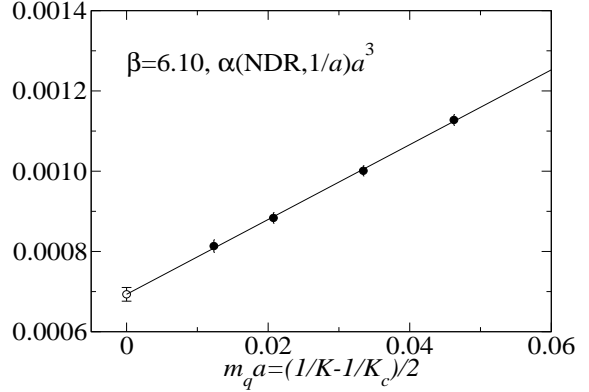}
\vspace{-8mm}
\caption{
Typical light quark mass dependence of the $\alpha$ parameter.
}
\vspace{-5mm}
  \label{fig:lqdep}
\end{figure}

\section{Simulation details}

Our calculation is carried out in quenched QCD
with the Wilson quark and the plaquette gauge actions.
The simulation parameters are given in Table\ref{simulation_parameters}.
We adopt three values of lattice spacings in the range
$a\approx$ 0.1$-$0.064 fm.
The spatial lattice size is about 3 fm
to avoid finite size effects.
We employ four quark masses corresponding 
to $m_{PS}/m_{V}\approx$ 0.75$-$0.5 
for each $\beta$.

For smearing the source, the quark wave function in the pion 
is measured for each hopping parameter 
on gauge configurations fixed to the
Coulomb gauge except for $t$=0 time slice 
where the wall source is placed\cite{wogf}.
To estimate $\sqrt{Z_p}$ we first fit the smeared-local proton correlator
to a single exponential form $Z_p^\prime {\rm exp}(-m_p t)$,
and then the local-local proton correlator is fitted 
to $Z_p {\rm exp}(-m_p t)$ with $m_p$ fixed to the value determined
by the smeared-local correlator.

The matrix elements defined on the lattice are converted  to those 
in the continuum at $\mu=1/a$
and the scale is evolved up to  $\mu$=2GeV using
the two-loop renormalization factor in the continuum\cite{Nihei:1994tx}.

\section{Results}

Figure~\ref{fig:lqdep} shows
the light quark mass dependence of the $\alpha$ parameter at $\beta=6.1$,
which is well described by a linear function.
Similar quark mass dependence is found for the $\beta$ parameter. 
We check that linear and quadratic extrapolations yield 
consistent results within error bars in the chiral limit at all the
lattice spacings.

\begin{figure}[t]
  \includegraphics[width=\minitwocolumn,height=5.2cm]{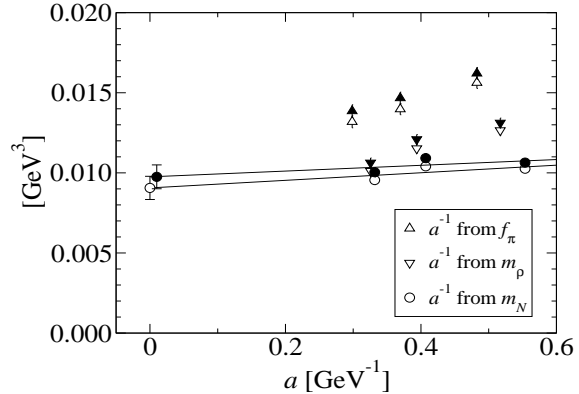}
\vspace{-8mm}
\caption{
$\alpha$ (open) and $\beta$ (filled) parameters 
as a function of lattice spacing.
}
\vspace{-5mm}
  \label{fig:adep}
\end{figure}

In Fig.~\ref{fig:adep}, we plot the parameters $\alpha$ and $\beta$ 
in physical unit as a function of lattice spacing $a$,
where we experiment with three choices $m_N$, $m_\rho$ and $f_\pi$
as physical input to determine the lattice spacing.
%
In case of the nucleon mass $m_N$ as physical input
we find little scaling violation, which allows us 
to take the continuum limit of the parameters 
by linear extrapolation.
We take the extrapolated value
as the central one of the $\alpha$ and $\beta$ parameters 
in the continuum limit.
On the other hand, the parameters show rather large scaling violation
if we set the lattice spacing by the rho meson mass $m_\rho$
or the pion decay constant $f_\pi$.
Although in these cases the values at the continuum limit 
possibly deviate from the result obtained with $m_N$ as physical input,
simple linear extrapolation is not reliable to
estimate the deviation. 
Instead, we estimate this ambiguity from the result of quenched 
light hadron mass spectrum obtained 
by the CP-PACS collaboration\cite{Aoki:2002fd}, 
which shows that
the values of $m_N$ and $f_\pi$ in quenched QCD deviate 
from the experiment by about 7\% and 10\% 
respectively in the continuum limit,
once we set the lattice spacing by $m_\rho$.

Taking account of this ambiguity, we obtain
\begin{eqnarray}
 |\alpha(2\textrm{GeV})|&=&0.0091(08)
 \left(^{+10}_{-19}\right)
 \textrm{GeV}^3,\\
 |\beta(2\textrm{GeV})|&=&0.0098(08)
 \left(^{+10}_{-20}\right)
 \textrm{GeV}^3,
\end{eqnarray}
with $\alpha/\beta<0$ in the continuum limit,
where the first error is statistical and the second one is
due to scale setting.  We note 
that the relative sign of $\alpha$ and $\beta$ could be important, whereas
the overall one is irrelevant.


\vspace*{5mm}
This work is supported by Large Scale Simulation Program
No.~98 (FY2003) of High Energy Accelerator Research Organization
(KEK), and also in part by the Grant-in-Aid of the Ministry of
Education (Nos. 12740133, 13135204, 13640259, 13640260, 14046202,
14540289, 14740173, 15204015, 15540251, 15540279).
N.T. is supported by the JSPS Research Fellowship.

\end{document}